# All-optical modulation with single photons using electron avalanche

Demid V. Sychev[1,2,3,4,*,§], Peigang Chen[1,2,3,4,§], Yuheng Chen[1,2,3,4], Morris Yang[1,2,3,4], Colton Fruhling[1,4], Alexei Lagutchev[1,4], Alexander V. Kildishev[1,2], Alexandra Boltasseva[1,2,3,4], Vladimir M. Shalaev[1,2,3,4*]

[1]*Birck Nanotechnology Center, Purdue University, West Lafayette, IN 47907, USA*

[2]*Elmore Familly School of Electrical and Computer Engineering, Purdue University, West Lafayette, IN 47907, USA*

[3]*Purdue Quantum Science and Engineering Institute, Purdue University, West Lafayette, IN 47907, USA*

[4]*Quantum Science Center, National Quantum Information Science Research Center of the U.S. Department of Energy, Oak Ridge, TN 37931, USA*

§ Equally contributed.

**Emails:* sychev@purdue.edu & shalaev@purdue.edu

**Abstract:** The distinctive characteristics of light, such as high-speed and low-loss propagation, low cross-talk and low power consumption, along with photons unique quantum properties, make it most suitable for various applications in communication, high-resolution imaging, optical computing, and emerging quantum information technologies. One limiting factor, though, is the weak optical nonlinearity of conventional media that poses challenges for the control of light with ultra-low intensities. In this work, we demonstrate all-optical modulation enabled by electron avalanche process in silicon, using a control beam with single-photon light intensities. The observed process corresponds to a record-high nonlinear refractive index of $n_2 \sim 1.3 \times 10^{-2}\,\mathrm{m}^2/W$, which is several orders of magnitude higher than the best known nonlinear optical materials. Our approach opens the possibility of gigahertz-speed, and potentially even faster, optical switching at the single-photon level, which could enable a family of novel on-chip photonic and quantum devices operating at room temperature.

**Main text:**

Realizing efficient all-optical modulation is of critical importance for many scientific explorations and technological applications[1–8]. The interaction between light beams consisting of a macroscopic number of photons is usually realized through nonlinear optical media with second- $\chi^{(2)}$ or third-order $\chi^{(3)}$ nonlinearities[9,10]. For example, the Kerr effect in $\chi^{(3)}$ materials results in a refractive index $n$ that depends on the intensity of the propagating light[9]. This dependence enables all-optical modulation and other optically-controlled functionalities[11–13]. However, for conventional materials, the nonlinear coefficients are usually small. Treated as a small perturbative effect, achieving appreciable nonlinear response thus requires relatively high intensities to accomplish all-optical modulation. Therefore, this effect is generally not suitable for applications at a single-photon intensity level.

To overcome the weak interaction of traditional nonlinear materials, optical cavities are often incorporated into experiments, which are then conducted in the so-called strong-coupling regime. For example, this was demonstrated with quantum emitters (QEs) coupled to photonic cavities[14–23]. Additionally, much progress has been achieved with photonic polaritons[24], where single-photon intensity switching was demonstrated at room temperature. Other systems with high nonlinearities include photonic avalanche systems[25] and electronic devices mimicking the optical nonlinear response[26]. However, most of these approaches rely on high-finesse cavities or slow electronics, which severely limits the bandwidth, the speed and the operational wavelength range. Many approaches are also implemented at cryogenic temperatures, limiting their practicality.

In this work, we present a new approach to the realization of single-photon switches at room temperature that is cavity-free and offers a broad spectral operational range. We propose and demonstrate the use of the avalanche multiplication process in a CMOS-compatible semiconductor[27–29] to significantly alter the refractive index of silicon. A pulse with single-photon intensities triggers the avalanche process, and the refractive index change is observed with a weak probe pulse (**Fig 1a**). Importantly, the use of standardized materials could allow for rapid commercial development and deployment of the proposed device concept.

The refractive index and absorption in silicon are influenced by factors such as the evolving free charge carrier concentration[30,31], and thermal effects[32,33]. The carrier concentration is affected as photons with energies greater than the band gap of the semiconductor are absorbed. The absorption promotes valence-band electrons to the conduction band and increases the concentration of free charge carriers. The refractive index change is described according to the Drude model via the following expression[30]

$$\Delta n = -\frac{\lambda^2 e^2}{8\pi^2 c^2 \varepsilon_0 n_0}\left(\frac{\Delta \mathrm{N}_e}{\mu_e^*} + \frac{\Delta \mathrm{N}_h}{\mu_h^*}\right), \qquad (1)$$

where $e$ is the elementary charge, $\varepsilon_0$ is the vacuum permittivity, $\lambda$ is the wavelength, $n_0$ is the unperturbed refractive index of silicon and $\mu_e^*$ ($\mu_h^*$) is the effective mass of electrons (holes). The newly promoted electrons are thermalized within the semiconductor increasing the temperature. This affects the refractive index in accord with the formula $\Delta n = \frac{\partial n}{\partial T}\Delta T$, where $\frac{\partial n}{\partial T} \cong 1.85 \times 10^{-4}\ K^{-1}$ [33]. Such changes in the optical properties have been used for both optical modulation, for example in self-electro-optical devices, and for read-out/sensing purposes,or in laser voltage probing[34,35].

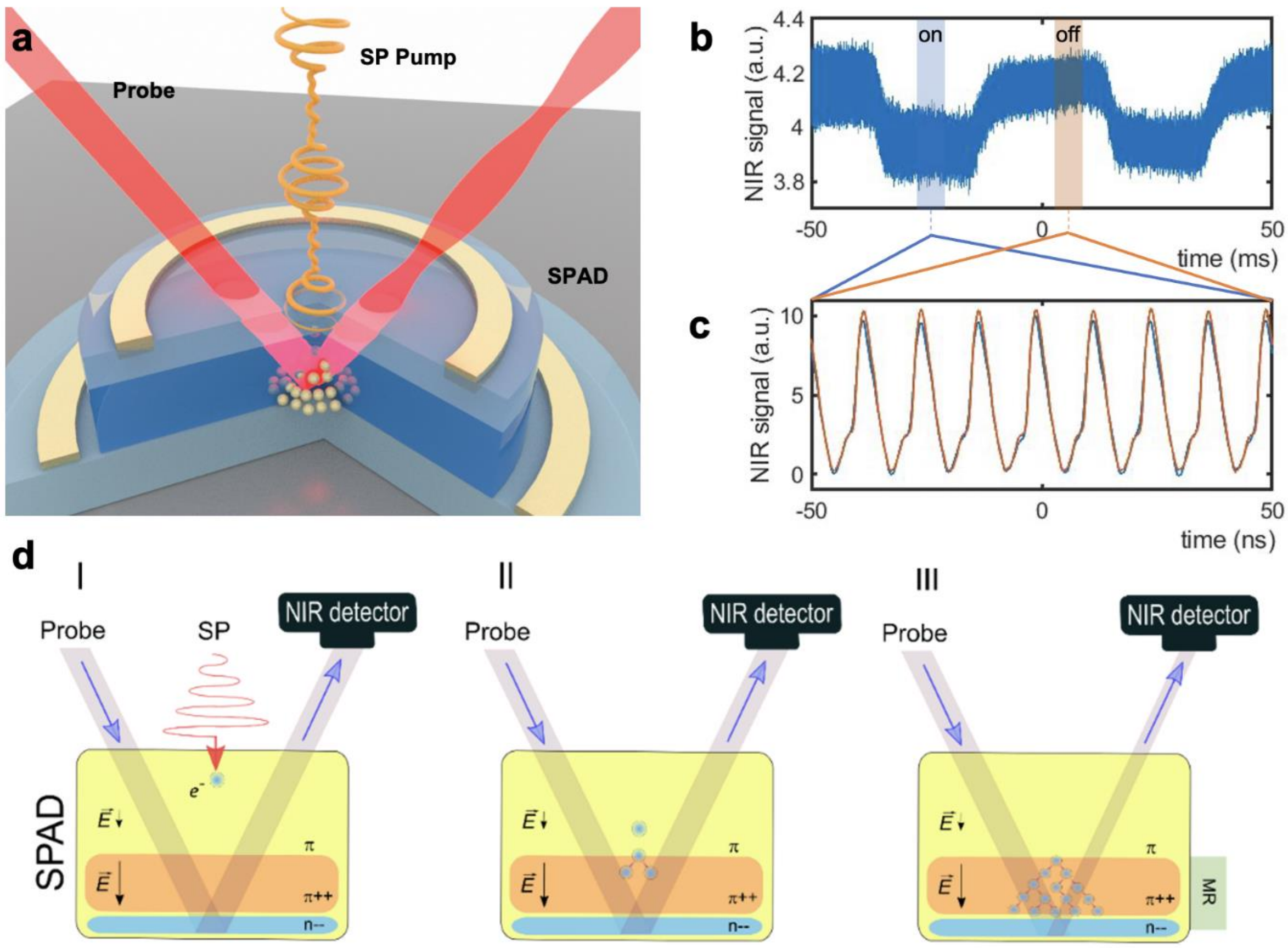

**Fig. 1: a, Schematics of the pump-probe experiment** with a single photon (SP) control (pump) beam modulated by a mechanical chopper. **b,** The NIR probe signal averaged over 100 data sets in the presence/absence (on/off) of the control beam at 810 nm wavelength, with about 0.06 photons per pulse at 20Hz chopping rate. **c,** A zoom-in on to the NIR probe signals at the nanosecond time scale averaged over 100 data sets as in **b**. The periodic structure reflects NIR laser pulses at 80 MHz. Areas for on/off states clearly show different amplitudes, proving the optical nature of the modulation effect. **d,** Principle of single-photon modulation: *(I)* A single photon of the control beam is absorbed and creates a single electron in a conduction band of a semiconductor. The electron is then accelerated towards the multiplication region (MR, p-n junction) by an externally applied electric field. *(II)* Once in the MR, the electron initiates an avalanche multiplication effect and injects more electrons into the conduction band. *(III)* An avalanche of electrons causes a significant change in the free charge carrier concentration and, subsequently, in the refractive index, altering/modulating the intensity of a reflected NIR probe beam.

In the laser voltage probing and other conventional methods of modulation through charge carrier injection, the number of injected carriers $\Delta N_e$ ($\Delta N_h$) is proportional to the number of absorbed photons. At single-photon intensities considered here, the effect of a single electron promotion on the refractive index is practically immeasurable. Therefore, in our approach, the carrier concentration is dramatically amplified using the electron avalanche effect, which is used in single-photon avalanche diodes (SPADs) [36] and other semiconductor devices[37]. The light-induced carrier concentration in this case is amplified by several orders of magnitude, specifically, by the avalanche multiplication factor $M$ that can be up to million[27–29,38]. Therefore, the absorption of a single photon results in a significant increase in the charge carrier density and thus substantially affects the refractive index and the absorption coefficient of the material.

Let us delve deeper into the avalanche process since it is important for understanding the dynamics discussed below. In SPADs, an avalanche occurs when the bias voltages are above the breakdown voltage in the so-called Geiger mode. Under these conditions, a photon-induced electron is first generated via optical absorption (**Fig. 1d** I). The electron is then accelerated by the bias voltage and triggers the creation of additional free electrons into the conduction band[27,28] (**Fig. 1d** II). This process, in turn, leads to a cascade of electron-electron collisions that results in an exponential growth of the charge concentration in the semiconductor (**Fig. 1d** III). This avalanche rise of free charge carriers typically occurs very fast, in the sub-nanosecond time scale. We refer to this process as a “fast” effect. Newly injected electrons possess relatively high energy, causing heating of the material around. Material’s heating via “hot” electrons that happens through the interaction with phonons is also fast, within the same sub-nanosecond time scale. Subsequent relaxation then depends on thermal conductivity and other factors such as the charge transport in the p-n junction area, and it is relatively slow, in the microsecond time scale. Therefore, we refer to this thermal effect as “slow.” In both cases, the light-induced change in the refractive index is strongly enhanced by the multiplication of charge careers caused by an avalanche and it can be detected optically by a probe beam in the near-infrared (NIR) wavelength range. NIR photons have lower energy than the bandgap of silicon, lowering the probability of electron injection compared to a visible-range control beam.

The exponential growth of charge carriers occurs in the region between the donor (n-doped) and acceptor (p-doped) regions of the semiconductor, where the fields are the strongest. This region is called the multiplication region (MR)[29,38]. When $N$ electrons enter this multiplication region, they generate $M \times N$ electrons at the output, where $M$ represents the multiplication factor. For SPADs, the $M$- factors of $10^5$-$10^6$ are routinely achieved. As a result, the initial photo-excited electron results in a change of the free charge carrier concentration of up to $1.2 \times 10^{18}$cm$^{-3}$, for $M{\sim}10^6$ (see **Supplementary Materials, S4**). This change of the carrier concentration corresponds to the refractive index change of $\Delta n \approx -1.1 \times 10^{-3}$, according to **Eq. (1)**. In this work, we demonstrate all-optical modulation with a beam of single-photon intensities using the effect caused by the avalanche multiplication process.

To investigate the optical modulation and its temporal response, we conducted pump-probe measurements with a variable time delay between the pulses. Two different sets of lasers are used to access different repetition rates for different measurements. First, we use a 100fs 80MHz pulsed laser system, with the pump at 810 nm wavelength and the probe pulse at 1550 nm. The pump at 810 nm wavelength from the laser (Spectra-physics MAI TAI HP) is sent into OPO (Radiantis ORIA IR) to generate the probe pulse at 1550 nm, where the pulse from Mai Tai is used as a pump/control pulse, while the output pulse from OPO is used as a probe pulse (for details, see **Supplementary Materials S1**). The average number of photons per pulse in the control beam is approximated as the ratio of the count rate to the laser repetition rate. The strongly attenuated pump pulses are utilized as an ultra-weak control beam to trigger an electron avalanche within the SPAD structure. The intensity of the reflected probe beam was measured using a standard InGaAs detector (Thorlabs PDA05CF2) and analyzed with either a lock-in amplifier or an oscilloscope. We implement a difference (on/off) detection scheme utilizing a mechanical chopper for the control beam synchronized with the lock-in amplifier **(Fig. 1b, c)**.

To acquire data at low repetition rates (200 kHz-7 MHz), we used a pair of pulsed lasers operating at 520 nm and 1550 nm wavelength, respectively, with pulse durations of <130 ps. The experiments in this case consist of a pump-probe setup with two light pulses incident on a commercial SPAD structure (PerkinElmer SPCM-AQR-15 or Excelitas SPCM-AQRH-10-ND). The light pulses are generated by a pair of pulsed lasers with wavelengths centered at 520 nm (pump) and 1550 nm (probe). These sources are triggered by an electronic pulse generator (Stanford Research DG645) with variable repetition rate and delay between pulses. The delay between the optical pulses is controlled with 5-ps precision, in a wide range from picoseconds to several microseconds. The SPAD structure can be approximated as a disk of 180µm in diameter with few tens of microns in thickness(**Supplementary Materials, S3**)

The control beam intensity was set such that the mean photon number per pulse was in the range $0.1 - 1$ photons. Approximating the pulsed laser as a coherent source, the number of photons per pulse of the control beam is described by the Poisson distribution. The probability of $m$ photons is $p_m = \mathrm{e}^{-\langle m \rangle} \frac{\langle m \rangle^m}{m!}$, where $\langle m \rangle$ is the mean number of photons per pulse. For $\langle m \rangle \sim 0.1$, the probability of detecting two photons $p_2$ is approximately 20 times lower compared to the probability of detecting a single photon $p_1$, which is effectively equivalent to the intensities of a source of single photons. For simplicity, we assume that the average number of photons $\langle m \rangle$ equals to the probability $p_{vis}$ of the control pulse to create an avalanche so that $\langle m \rangle = 1$ causes the avalanche with 100% probability.

As mentioned, in a typical SPAD, the avalanche develops on sub-nanosecond timescales[39], which defines its rise time. In contrast, the recovery time (dead time) for SPAD is governed by the speed of the quenching circuit and is typically on the order of several tens of nanoseconds[36]. We, therefore, scan the pump-probe delay from -40ns to +40ns to observe the relevant dynamics. We plot the relative difference of NIR beam reflection between "on" and "off" states initiated by the pump/control beam, with the average number of photons per pulse $\langle m \rangle = 1$, for the control beam, and 30mW power, for the probe beam. The results reveal a strong dependence of the observed probe beam modulation on the delay between the control and the probe pulses (**Fig. 2a**). The plot displays a distinct gap with a near-zero signal in the delay range of approximately from -17ns to 17ns, flanked by two slow-decaying regions, which are symmetrically positioned around the plot's origin, with the two peaks on each side. The ±17ns region near the zero can be attributed to the deadtime of the SPAD, which is 16 ns according to the datasheet for the SPAD we used (see **Supplementary Materials, S5**).

The signal at larger negative delays, when the probe pulse arrives before the pump pulse (region (i) in Fig. 2a), is attributed to a process with the relaxation time exceeding the deadtime. This is the "slow" effect discussed above, with relatively slow relaxation time. To explore the timescale of relaxation for this "slow" effect, we extended the pump-probe delay range to 0–1800 ns (**Fig. 2b**) conducted at 500kHz repetition rate. Over this extended range, the curve shows a gradual decay with the characteristic time constant of approximately 13 µs, as determined by the fitting.

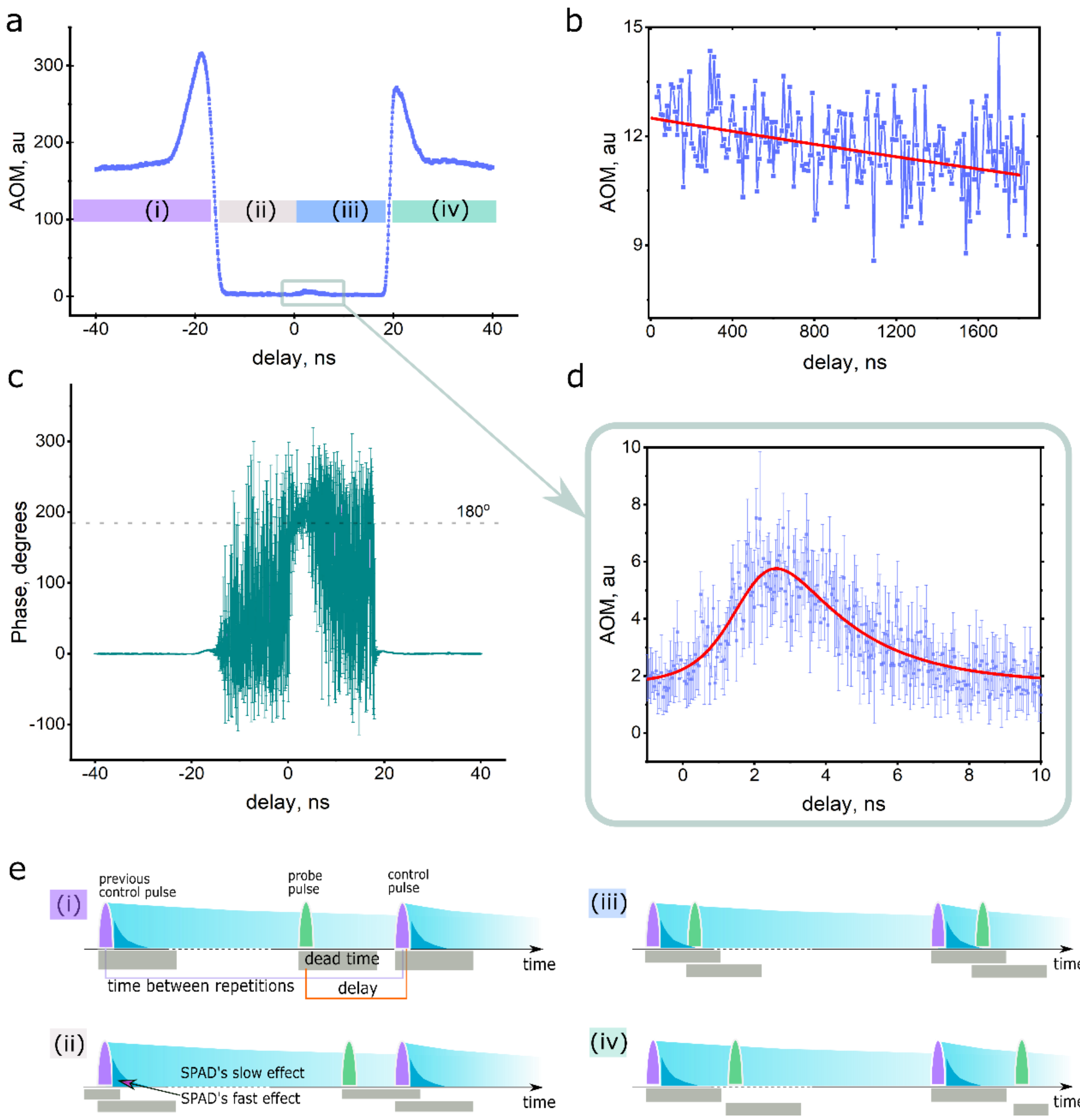

**Fig. 2: Time dependence of modulation. a,** The amplitude of NIR beam modulation (all-optical modulation, AOM) as a function of the time delay between the control beam and probe pulses. **a,** The amplitude of NIR modulation on a bigger time scales for the delay. The red line is a fitting function. **c,** The relative phase difference between the modulation of NIR probe beam and a chopper's reference signal corresponding to the plot in figure **a**. **d,** The amplitude of NIR modulation on a smaller time scale. **e,** Time diagram of pulse sequence for each case of time delay between control and probe pulses.

This decay time is about 65 times slower than the repetition rate (5 MHz) used for the measurements in **Fig. 2a**. Such a slow response time aligns with the timescales typically associated with thermal effects in silicon structures[40,41]. Importantly, the rise time for this "slow" effect is fast, on the nanosecond time scale, which can be seen from the immediate rise of the signal at around 17 ns delay in **Fig.2a**. This feature makes the considered "slow" effect significantly different from conventional thermal all-optical modulators where both time scales, for the rise and for the decay, are at the microsecond time scale[40,41]. The presence of the two peaks with a higher modulation amplitude could be attributed to the SPAD control circuit, which applies a higher voltage immediately after the SPAD is reset. Further details related to data shown in **Fig. 2a** can be found in **Supplementary Materials, S5.**

Next, we investigate the “fast” dynamics of the observed process (see “fast” region of **Fig. 2a**). Note that the near-zero range in Fig. 2a is symmetric with respect to the pump-probe zero-delay. This occurs because at high enough powers the NIR probe beam starts to induce an avalanche in the SPAD due to the two-photon process and/or the presence of mid-gap defects. Thus, regardless of which pulse (the control/pump or the probe) initiated the avalanche, it does not allow the subsequent pulse to create another avalanche. This effectively results in the probe detected signal being the same for control “on” and “off” states. This means that the amplitude of the AOM signal, which is given by the difference in the detected probe for the control being in the “on” and “off” states, goes to zero in both negative and positive time delays around zero (within the dead time). Our findings confirm that the absence of the AOM signal lasts for about 34ns (from -17ns to 17ns), which roughly corresponds to twice of the dead time of the SPAD used in our measurements.

We also observe a small peak on the positive side of the near-zero delay corresponding to the fast effect (**Fig. 2d**). The fit shows the time constants of about 1ns and 2ns, for the rise and the relaxation processes, respectively. The modulation amplitude for this peak is about 40 times smaller compared to the “slow” effect. Note that we can see this smaller fast effect because of the disappearance of the “slow” AOM effect at near-zero delays, when both the pump and the probe can cause the avalanche. Interestingly, the “fast” and “slow” responses show the opposite signs of the modulation effect. This can be clearly seen from the phase difference measurements between the modulated NIR probe beam and the chopper reference signal in the lock-in amplifier (see **Fig. 2c)**. This alludes to the different physical phenomena responsible for each modulation. The “fast” effects cause a decrease in the probes reflected amplitude (180° in **Fig. 2c**), while the “slow” effect results in an increase (0° in **Fig. 2c**).

Next, we study the sensitivity of modulation to the probe and control beams intensities (**Fig. 3**). We measured the response with the pump-probe delay set to 2ns and 22ns to investigate the “fast” and “slow” regimes separately. First, we measure the dependence of the modulation amplitude of the NIR probe beam as a function of the NIR probe power. As before, the intensity of the control beam is chosen to maximize the probability of an avalanche, with $\langle m \rangle = 1$ per pulse corresponding to $p_{vis} = 1$. In the case of a “slow” regime at 22ns delay, the curve in Fig. 3c, as expected, demonstrates a monotonic linear growth for the reflectivity modulation with the increase in the probe power. Interestingly, for the “fast” regime (2ns delay), the curve experiences non-monotonic behavior (**Fig. 3a**). Below 15mW for the probe power, the AOM decreases linearly with the NIR power. This is because for a low-intensity probe, the avalanche is caused only by the control beam so that the AOM is at its maximum and it is entirely due to the “slow“ effect, which is 40 times larger than the fast effect. However, when the probe power increases it also starts causing the avalanche so that the AOM signal (as mentioned, it is given by the difference in the detected probe signal for the control beam in “on” and “off” states) drops. At 15mW probe beam power, the probability that the NIR also causes the avalanche approaches unity so that the signal difference between the control/pump “on” and “off” - and thus the AOM - is near zero. With the further increase of the probe intensity, the fast modulation starts to dominate, and it keeps growing with the increase of the NIR power, eventually surpassing its initial value. For further details, see **Supplementary Materials, S5.**

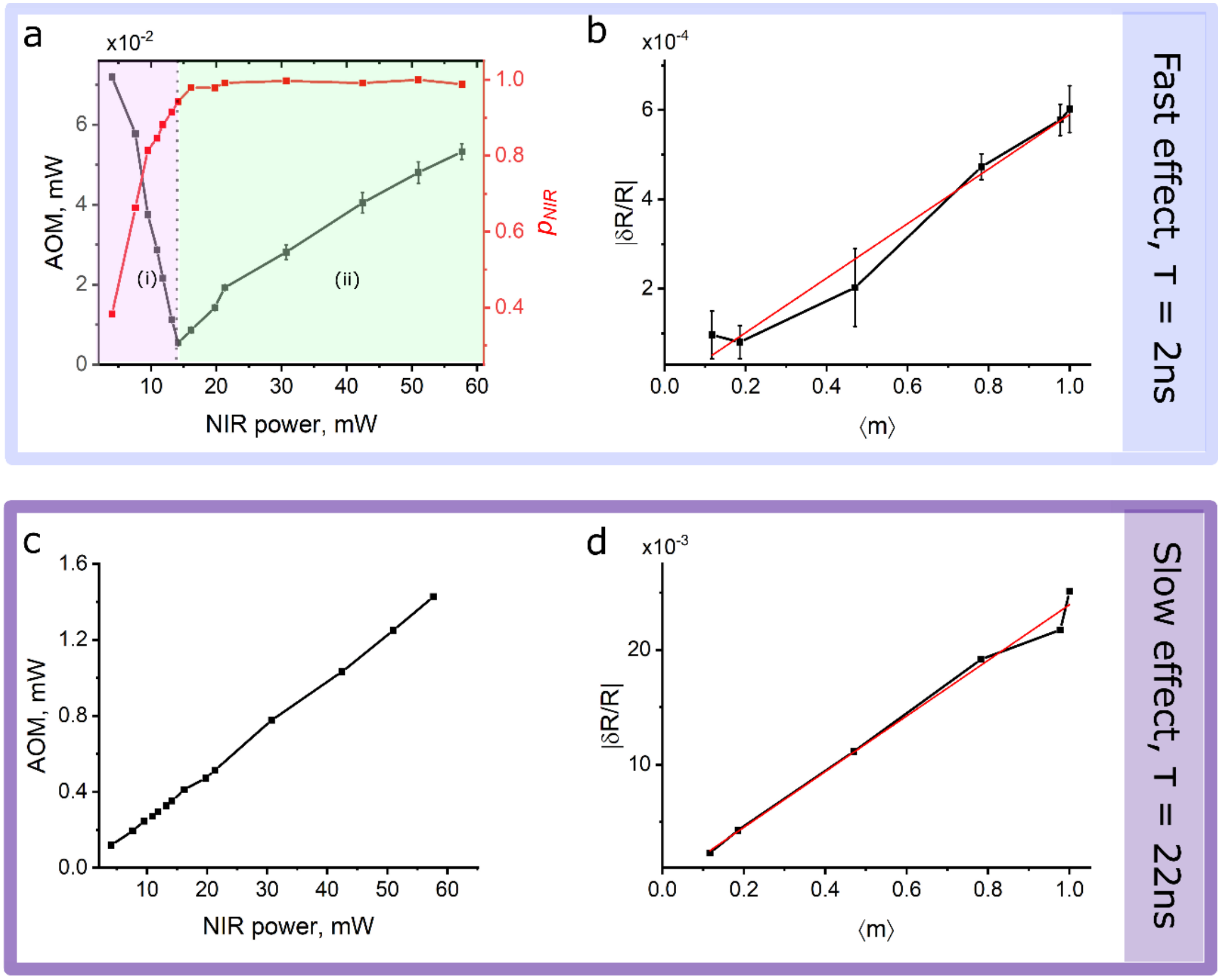

**Fig. 3: Dependencies of a signal on intensities of the control and probe beams. a,** the amplitude of NIR probe beam modulation (all-optical modulation, AOM) as a function the probe's beam intensity for 2ns delay between the control and probe beams. The red line indicates the probability of avalanche induced by a NIR pulse $p_{NIR}$ as a function of NIR beam power. **b,** Relative modulation of the reflected NIR signal as a function of the intensity of the control beam at 2ns delay between the control and probe pulses. The control beam's intensity is shown in units of average number of photons $\langle m \rangle$ in a single pulse. In figures **c** and **d** the amplitude of the NIR probe modulation is shown as a function of NIR probe and control beam intensities, respectively, for 22ns delay between pulses. The red lines in figures **b** and **d** are the fitting functions.

To extract the magnitude of the nonlinearity, we measure the modulation amplitude of the NIR probe beam as a function of the control beam intensity. For both the "fast" and "slow" effects it demonstrates linear growth with the increasing number of photons $\langle m \rangle$ in the control beam pulse (**Fig. 3b, d**). The NIR beam's intensity in these measurements is chosen to be 30mW.

The refractive index modulation $\Delta n$ is evaluated through the relative modulation of the reflectivity $\Delta R/R$. Generally, the reflection from a structure depends on the combination of various effects, including internal interference, the absorption coefficient, and the refractive index. Here, we assume that the change in refractive index is the dominant effect, which is justified for obtaining a rough estimate. Then $\Delta n$ can be evaluated using the standard Fresnel formulas for reflectivity, which, to the first order of approximation, gives $\Delta n \cong$ $-\frac{1}{4}(n_0^2-1)\Delta R/R$, where $n_0$ is the refractive index of silicon, $R$ is the reflectivity, and $\Delta R$ is the amplitude of the reflectivity modulation.

Using this formula, we find the refractive index modulation for the "fast" and "slow" responses to be $\sim -1.7\times10^{-3}$ and $\sim 7\times10^{-2}$, respectively. According to **Eqn. (1)**, the $\Delta n$ from the "fast" effect corresponds to a free charge carrier concentration modulation of $\sim 2.5\times10^{18}$cm$^{-3}$, which is consistent with the concentration estimated above. By analogy to the Kerr optical nonlinearity, the change of the refractive index is proportional to the intensity of the beam $\delta n = n_2 I$, which can be used as an approximation, for the small number of photons used in the experiment. The important difference with the conventional optical Kerr effect is that the electron avalanche shows a significant index change at much lower intensities down to the level of single-photon intensities. The estimate for $n_2$ coefficient gives values around $-3.3\times10^{-4}$ m$^2$W$^{-1}$ and $1.3\times10^{-2}$ m$^2$W$^{-1}$ for the "fast" and "slow" responses, respectively, which both are orders of magnitude higher compared to $n_2$ values of commonly used nonlinear optical materials (see **Table S1** in **Supplementary Material**). For a detailed estimate of the sensitivity of the avalanche based all-optical modulation, see **Supplementary Materials**, **S7**.

**Discussion and outlook**

We demonstrated the modulation of NIR light by a visible-wavelength pulse with an average number of photons $\langle m\rangle \sim 0.1-1$ per pulse, using the avalanche multiplication process in commercially available SPAD system. The observed effect is caused by the light-induced refractive index changes, which have different physical nature. The two underlying phenomena exhibit different time characteristics and lead to the index change with different magnitudes and signs. The first effect rises on the nanosecond time scale and then decays relatively slowly, on the order of microseconds, while the second one rises and decays on the nanosecond time scale. According to these characteristics, the first, slow-decaying process can be attributed to thermal modulation, while the second one is due to free carrier dynamics. Both effects are amplified by the avalanche multiplication process, which makes them sensitive to even single-photon intensities. In these avalanche-amplified processes, changes of the refractive index of $-1.7\times10^{-3}$ and $7\times10^{-2}$ have been measured, which are induced by the control beam with single photon intensities, for the "fast" and "slow" phenomena, respectively. These strongly enhanced optical responses are equivalent to the nonlinear optical coefficients $(n_2)$ of $-3.3\times10^{-4}$ m$^2$W$^{-1}$ and $1.3\times10^{-2}$ m$^2$W$^{-1}$, where the highest number, for example, is seventeen orders of magnitude larger compared to that in lithium niobate and more than fifteen orders of magnitude compared to silicon (see table S1 in **Supplementary Materials**).

The proposed approach could offer significant advantages for several practical applications. Importantly, this modulation works at room temperature and offers CMOS compatibility with the possibility for on-chip integration[42]. In addition, it operates within a broad wavelength range (from 400 nm to 1000 nm for silicon SPAD). The proposed technique shows great potential for achieving extremely strong optical nonlinearities at single-photon intensities, which paves the way for a variety of different applications[43–45]. The method can be also extended further for application in electrooptical modulation. Compared to conventional methods[46–49] the avalanche-based modulator can provide a compact, on-chip solution with comparable time characteristics, directly paving the way for scalable programable photonic circuits.

Our results demonstrate that photo-induced electron avalanches can be employed to achieve all-optical modulation. There are several avenues to further optimize this design, both in terms of the optical setup and device implementation. The use of longer wavelengths can reduce the dark noise and improve shot-noise performance for read-out purposes. Additionally, the introduction of photonic cavities can enhance the efficiency and sensitivity of the system[50]. Considerable improvements can also be made to the SPAD design for optical modulation. For instance, since the optical readout relies solely on the local current, the presence and detection of the external SPAD current in the circuitry becomes unnecessary, which potentially can make this approach significantly faster (**Supplementary Materials, S6**). This also opens possibilities for enhancing avalanche parameters by incorporating absorption layers and optimizing voltage parameters within the structure.

Furthermore, alternative materials can be explored for improved performance. For instance, using GaAs or transparent conducting oxides instead of silicon may enable the faster dynamic of the avalanche process due to its lower effective mass, higher charge carrier mobility, and direct bandgap transition between the valence and conduction bands[38,51]. These potential avenues for improvement offer promising routes for advancing the capabilities and performance of the demonstrated system. We also note that the same approach could be extended to infrared (IR) signals by employing electron avalanches in IR photodiodes, such as antimonide-based avalanche photodetectors[52].

**Acknowledgments:** We thank Dr. Soham Saha, Dr. Alex Senichev, Professor Mordechai Segev, Professor Andrew M. Weiner, Mariam Gigauri for valuable comments.

**Author contributions:**
D.S., A. L. conceived and planned the experiments. D.S., P. C., M. Y., Y. C. and C. F. performed the optical measurements. D. S., P. C. and A. L. performed the analysis of experimental data. A. V. K. performed FEM simulations. D.S. and P. C. wrote the manuscript with support from A. L., A. V. K., A. B., and V.M.S. and contributions from all coauthors. A.B. and V.M.S. led the project. All authors discussed the results and commented on the manuscript. D.S. and P. C. equally contributed to the work.

**Funding:**
This work is supported by the U.S. Department of Energy (DOE), Office of Science through the Quantum Science Center (QSC), a National Quantum Information Science Research Center and by a collaborative project with DEVCOM Army Research Lab "Ultrafast Space-Time Photonics and Single-Photon Optical Modulators".

**Competing interests:** The authors are inventors on a provisional patent application related to this work filed by the Purdue Research Foundation (no. 63/461,564, filed April 24, 2023). The authors declare that they have no other competing interests.

**Data and materials availability:** All data are available in the main text and/or the supplementary materials.

# Supplementary Materials for

## All-optical modulation with single photons using electron avalanche

Demid V. Sychev[1,2,3,4,*,§], Peigang Chen[1,2,3,4,§], Yuheng Chen[1,2,3,4], Morris Yang[1,2,3,4], Colton Fruhling[1,4], Alexei Lagutchev[1,4], Alexander V. Kildishev[1,2], Alexandra Boltasseva[1,2,3,4], Vladimir M. Shalaev [1,2,3,4*]

[1]*Birck Nanotechnology Center, Purdue University, West Lafayette, IN 47907, USA*

[2]*Elmore Familly School of Electrical and Computer Engineering, Purdue University, West Lafayette, IN 47907, USA*

[3]*Purdue Quantum Science and Engineering Institute, Purdue University, West Lafayette, IN 47907, USA*

[4]*Quantum Science Center, National Quantum Information Science Research Center of the U.S. Department of Energy, Oak Ridge, TN 37931, USA*

§ Equally contributed.

**Emails:* sychev@purdue.edu & shalaev@purdue.edu

## S1.Experimental setup

The experiment is performed using a pump-probe configuration. Two different sets of lasers are used to access different repetition rates for different measurements. For the first approach, we use a 100fs 80MHz pulsed laser system (**Fig. S1**). The pump at 810 nm wavelength from the laser (Spectra-physics MAI TAI HP) is sent into OPO (Radiantis ORIA IR) to generate the probe pulse at 1550 nm, where the pulse from Mai Tai is used as a pump/control pulse, while the output pulse from OPO is used as a probe pulse.

To acquire data at low repetition rates (200 kHz-7 MHz), we use another pair of pulsed lasers (Thorlabs GSL52A and GSL155A) operating at 520 nm and 1550 nm, respectively, with pulse durations of <130 ps. To increase the output power for the 1550 nm NIR laser, the output beam is directed through an optical amplifier (Thorlabs EDFA100S). Both lasers are synchronized by triggering optical pulses with electron pulses from a Stanford Research DG645, which can vary the electron delay in the range of tens of nanoseconds with the accuracy of 5 ps.

Next, control (pump) and probe beams are separated by a polarizing beam splitter PBS1 (Thorlabs PBS204); filters $f_1$ (Thorlabs FBH1550-12) and $f_2$ (FESH0900) are used to block the residual wavelengths in the corresponding arms of the interferometer. The pump/control beam in the visible wavelength range in the upper arm (**Fig. S1a**) of the interferometer is then modulated by a mechanical chopper (Thorlabs MC2000B) in a frequency range from 20Hz to 1000Hz. Half-waveplates combined with a polarizing beam-splitter are installed to control the beam power in both arms (not shown in the Figure). In addition, to obtain a single-photon level of the pump intensity, it is equipped with a set of neutral density filters. The arm of the probe beam is equipped with a tunable delay to control the temporal separation between the pump and the probe pulses. Two beams are merged again by $PBS_2$ and are sent into an optical filter consisting of two lenses ($L_1$ and $L_2$, 50mm focal length) and a pinhole ($P_1$, 50$\mu$m diameter). An optical filter is used to clean the spatial modes of the beams and to match the pump (control) and the probe beams in space. After this, both beams are focused on the SPAD structure (PerkinElmer SPCM-AQR-15 or Excelitas SPCM-AQRH-10-ND) through an aspheric lens with NA of 0.5 ($L_3$, C240TMD-C). Both elements are mounted on manual xyz translational stages. The probe beam reflected from the structure is being collected into an InGaAs amplified photodetector (Thorlabs PDA05CF2) via 90:10 (R:T) BS (Thorlabs CCM1-BS015). Two 1500-nm-wavelength long-pass filters $f_3$ (Thorlabs FELH1500 x2) before the NIR detector are used to block the visible wavelengths in the reflected beam. To protect the SPAD from stray light, the entire setup is covered with a tissue impenetrable for light during the experiment. The output of the InGaAs detector is directly measured with a 4GHz oscilloscope (LeCroy waveMaster 804Zi) or with a 100kHz lock-in amplifier (SRS SR810). A 3.5GHz frequency counter (BK PRECISION 1856D, not shown in the figure) is used to monitor SPAD clicks. The SPAD structure has a circular active area around 180 $\mu\text{m}$ in diameter and has 55% photon detection efficiency at 810 nm wavelength[1]. The amplitude of the modulation is extracted directly from lock-in amplifier readings multiplied by a factor of $\frac{4}{\pi}$. By this factor, the amplitude of the rectangular square function is smaller compared to the amplitude of the sine wave, which approximates it to the first order of the Fourier series.

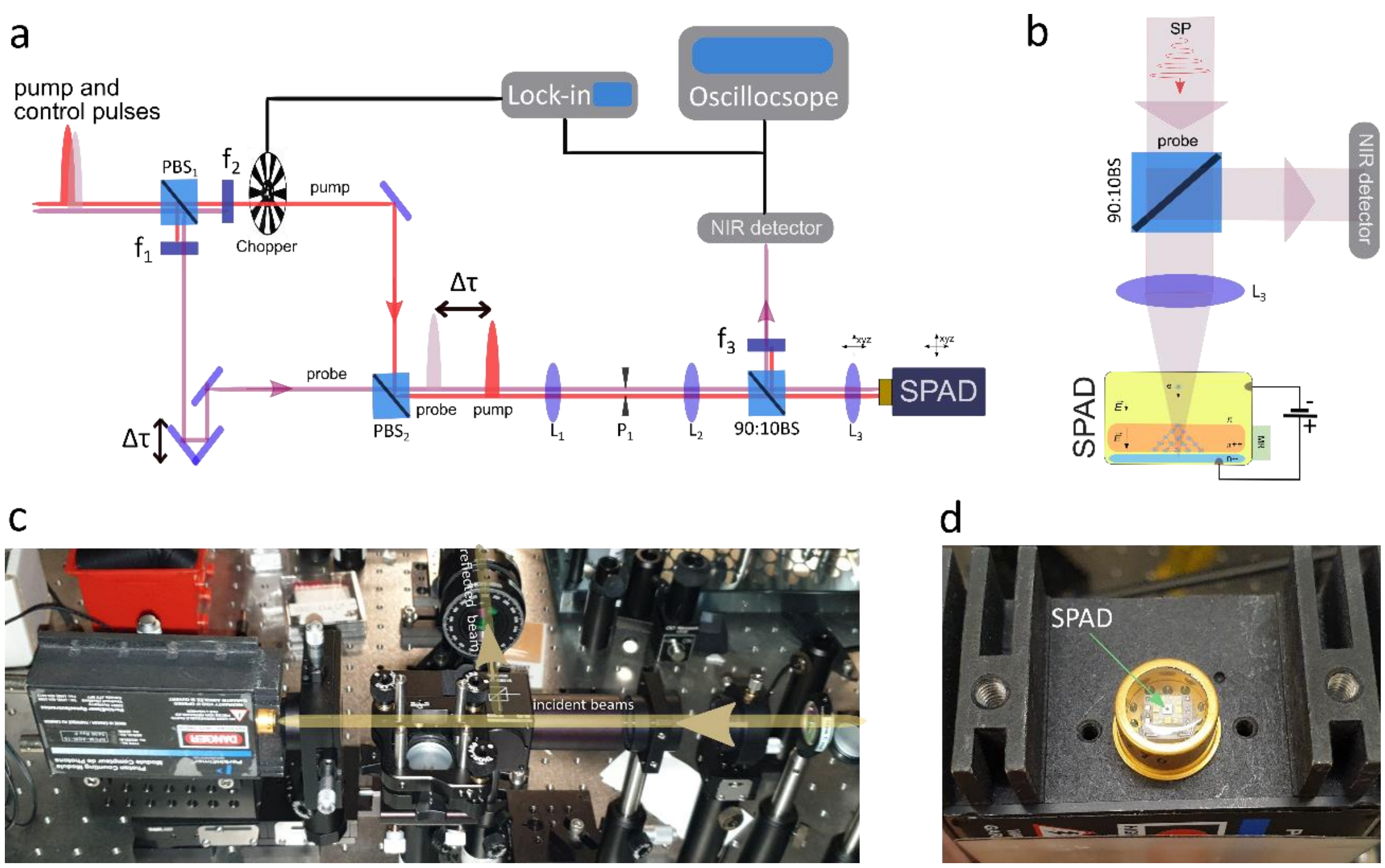

**Fig. S1: Experimental setup. a,** Schematics of the pump-probe experiment. **b,** The SPAD structure under the focusing lens. **c,** Part of the experimental setup. **d,** Photo of the SPAD structure used in the experiment.

## S2.Evaluation of $n_2$

For the Kerr effect, the refractive index depends on the beam intensity $I$ according to $n = n_0 + n_2 I$, where $n_2$ is a nonlinear coefficient of the optical Kerr effect. In our experiment, $n_2$ is evaluated from the change of the refractive index in the following way, $n_2 = \frac{n-n_0}{I} = \frac{\delta n}{E_{\mathrm{ph}} \frac{CR}{S}}$, where $E_{\mathrm{ph}} = h\nu$ is the energy of a single photon, and $CR$ is the count rate of SPAD, corresponding to the number of detected photons and the pump beam area $S = \pi \left(\frac{\lambda}{\pi\, NA}\right)^2$, defined after the focusing lens with a numerical aperture ($NA$) of 0.5 at a wavelength ($\lambda$) of 520 nm.

## S3.SPAD characterization

We approximate a typical SPAD design with a disk having the diameter of 180 $\mu\mathrm{m}$, as indicated in datasheets. The two-layer structure of the disk includes a multiplication region (approximately 500-nm-thick layer), which changes its refractive index when radiated by the control beam, and a 20-100-$\mu\mathrm{m}$ layer of doped silicon crystal (PerkinElmer SPCM-AQR-15). To evaluate the thickness of the SPAD structure, we measure the intensity of the reflected probe and its modulation as a function of the distance between the focusing lens and the SPAD structure (**Fig. S2**). From the comparison between the experiment and the simple model explained below, we estimate the thickness to be around 40 $\mu\mathrm{m}$. Moreover, the modeling explains some deviations in the shape of the curves, which are observed due to slightly different paths for the propagation of the two reflected beams, where one is reflected from the surface of the SPAD, while another one penetrates inside and is then reflected from the bottom of the structure (similar to the case described by the ellipsometric equation).

To simulate the operation of the system, we calculate the dependence of the power delivered to the InGaAs detector on the distance between the focusing lens and the SPAD (**Fig. S2a**). Such dependence occurs because the beam diameter is very sensitive to the distance between the focusing lens and the SPAD. To calculate the size of the output beam, we use the ABCD ray transfer analysis in the frame of the geometrical optics. Our model consists of three lenses, an InGaAs photodiode, and a SPAD structure modeled as a two-layer silicon structure with an absorption layer and a multiplication region where modulation of the refractive index occurs. Instead of modeling the reflection from the SPAD, we mirrored the SPAD, reducing the optical system to a one-way propagating geometry.

In the ray transfer method (ABCD method), the beam is presented as a two-dimensional vector where the first element is a distance from the optical axis, and the second one is the angle between the optical axis and the beam. Each of the elements in our model modifies the vector, which is done through subsequent matrix-vector multiplication. We used three types of 2×2 matrices, $R(n_1, n_2)$, $P(d)$, $L(F)$ corresponding to refraction, propagation, and thin lens transformation of a beam as a function of refractive indices $n_1$, $n_2$, the distance of propagation $d$, and the focal distance of a lens $F$. A more detailed description of the method with the definition of matrices can be found elsewhere[2].

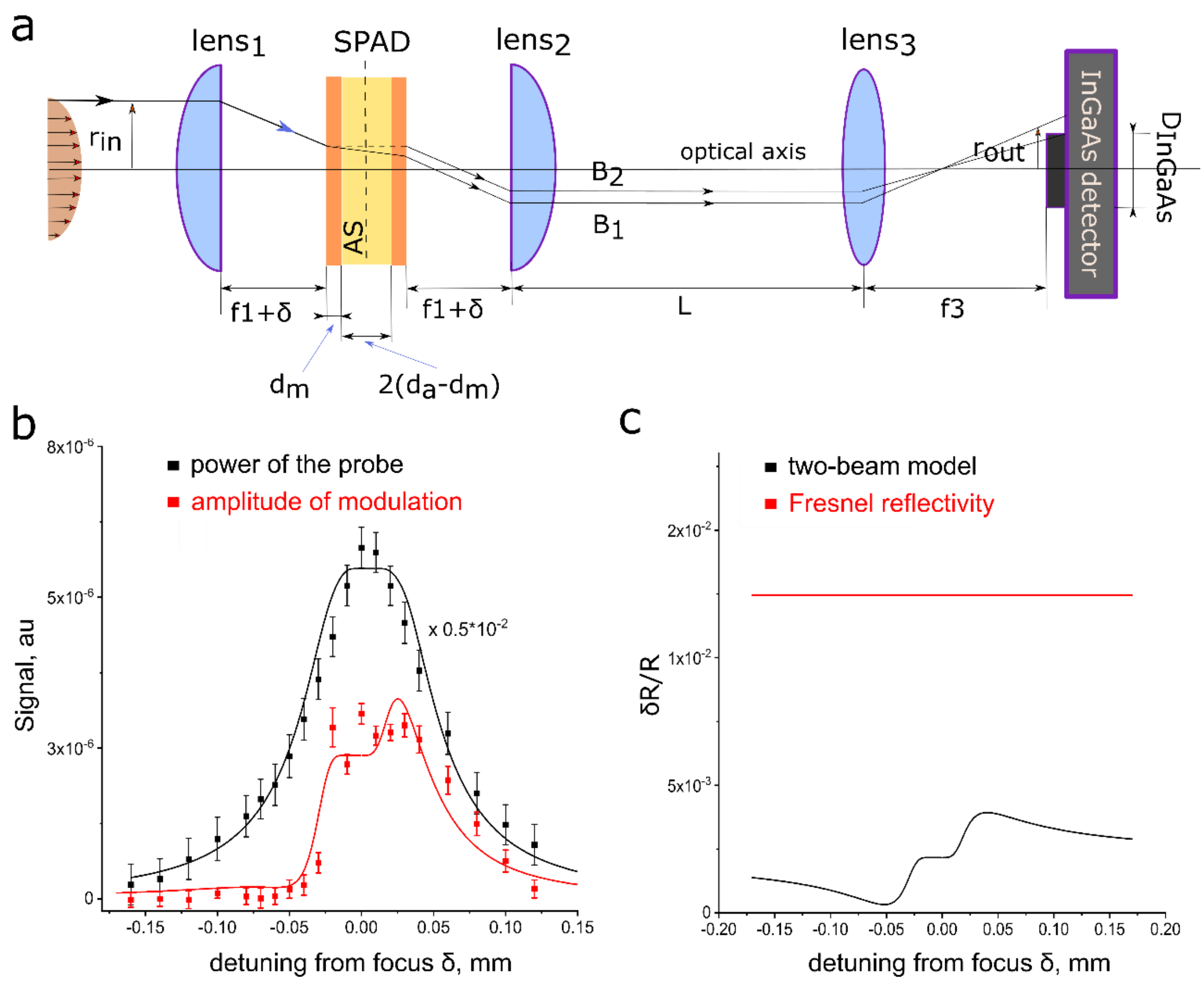

**Fig. S2: Simulation of ray propagation in the experimental setup. a,** Simplified schematics of the experimental setup. AS – axis of symmetry for SPAD, $B_2$ and $B_1$ indicate ray trace for reflected and deflected beams respectively. **b,** Comparison between the NIR amplitude of the modulation and the NIR reflected power as a function of the deviation from the focal point. Solid lines are the fitting functions. **c,** Relative reflectivity predicted by different models.

In the applied model, a collimated probe beam $B_{in} = \binom{r_{\text{in}}}{0}$ of a diameter $2r_{in}$ enters the optical system and the output beam, $B_{\text{out}} = \binom{r_{\text{out}}}{\theta_{\text{out}}}$ leaves the system. The radius of the output beam $r_{\text{out}}$ before the InGaAs detector is calculated for the output beam $B_{\text{out}}$. The final size of $B_{\text{out}}$ is calculated from,

$$B_{\text{out}} = P(f_3) \cdot L(f_3) \cdot P(L) \cdot L(f_2) \cdot P(f_1 + \delta) \cdot R(n_0 + \delta n, 1) \cdot P(d_m) \cdot R(n_0, n_0 + \delta n) \cdot P(2d_a) \cdot R(n_0 + \delta n, n_0) \cdot P(d_m) \cdot R(1, n_0 + \delta n) \cdot P(f_1 + \delta) \cdot L(f_1) \cdot B_{\text{in}},$$

where $n_0 = 3.5$ is an undisturbed refractive index of silicon; $\delta n$ – the amplitude of modulation of the refractive index in a multiplication region of a SPAD; $\delta$ – the deviation of the distance from the focusing distance of the lens and the SPAD from the focal length $f_1$; $f_{1,2,3}$ are focal lengths for lens$_{1,2,3}$ (8 mm, 8 mm, and 50 mm, respectively); $d_m = 500\text{nm}$, $d_a = 40\mu m$ are the thicknesses of the multiplication region and the absorption layer of the SPAD; $L = 1\text{m}$ is the distance between lenses 2 and 3 (**Fig. S2a**).

The diameter $2r_{out}$ of the output beam ($B_1$ or $B_2$) regulates the portion of light $S_{1;2}$ captured by the photodiode. For InGaAs detector with aperture $D_{\mathrm{InGaAs}}$ it can be evaluated as

$$S_{1;2} = PoL_{\mathrm{det}}(r_{\mathrm{out}}, r) = \frac{\sqrt{2\pi}}{r_{\mathrm{out}}}\int_0^{D_{\mathrm{InGaAs}}/2} \mathrm{e}^{-\frac{x^2}{2r_{\mathrm{out}}^2}} x\,\mathrm{d}x = 1 - \mathrm{e}^{-\frac{\left(\frac{D_{\mathrm{InGaAs}}}{2}\right)^2}{2r_{\mathrm{out}}^2}},$$

where the beam is assumed to be Gaussian.

When the probe beam hits the SPAD, it splits into two beams – reflected $B_1$ and deflected $B_2$. These portions of the beams can be estimated through the Fresnel equations, which give the corresponding coefficients. Thus, the portion of light $S_{\mathrm{probe}}$ of the probe beam as a function of the distance $\delta$ is given by

$$S_{\mathrm{probe}} = R\left[4\frac{n_0}{(n_0+1)^2}\right]^4 S_2 + \left(\frac{n_0-1}{n_0+1}\right)^2 S_1,$$

where $R = 0.4$ (from the fitting) is a coefficient of reflectivity (which is equivalent to the transmission through the axis of symmetry in our geometry) from the bottom of the SPAD.

Finally, to simulate how the amplitude of the modulation $S_{\mathrm{ampl}}$ depends on $\delta$ we calculate the difference in signals between the probe beam for refractive indices $n_0$ and $n_0 + \delta n$ so that

$$S_{\mathrm{ampl}} = S_{\mathrm{probe}}(n_0 + \delta n) - S_{\mathrm{probe}}(n_0).$$

Despite the simplicity of the presented model, it predicts the measured features very well. Functions $S_{\mathrm{probe}}$ , $S_{\mathrm{ampl}}$ are in good qualitative agreement with experimental data points (**Fig. S2b**). The asymmetrical shape of the curve for the amplitude is due to the superimposing of two slightly shifted curves for the reflected and deflected beams.

Notably, this model predicts approximately three times smaller relative reflectivity modulation $\frac{\delta R}{R} = \frac{S_{\mathrm{ampl}}}{S_{\mathrm{probe}}}$ compared to the reflectivity modulation from the Fresnel equation $\frac{\delta R}{R} = \frac{4\delta n}{n_0^2-1}$ for the same index modulation $\delta n$ (**Fig. S2c**). This occurs because the change in the refractive index in the two-beam model redistributes the probe power between the reflected and deflected rays, which still both hit the detector, unlike the model based for the Fresnel reflectivity.

## S4.Estimation of free charge carriers' concentration from a single event

The concentration of electrons can be estimated as $\frac{M}{d\pi D^2}$ (see **Fig. S3**), where $M \sim 10^6$ is the multiplication coefficient, which gives the number of electrons generated in the avalanche process by the initial electron, $d$ and $D$ are the dimensions of the spreading region of the electron cloud (see Fig. S2). While traveling through the multiplication region, the total number of electrons (thus their concentration) grows exponentially. The highest concentration is achieved near the end of the multiplication region. Assuming the length of the multiplication region is $l = 500$nm, which is typical for such structures[3–5], we can estimate the breeding coefficient $\alpha$ in the expression $M = e^{\alpha l}$. We consider the case when the number of electrons equals $\frac{M}{3}$, to be suitable for optical detection. Using this assumption, we estimate $d$ from the expression $\frac{M}{3} = \mathrm{e}^{\alpha(l-d)}$, which gives $d = \frac{3\ln 3}{\ln M} \sim 50$nm in this case. The lateral dimension can be extracted from the assumption of thermal diffusion of electrons, which gives $D = V_{\mathrm{th}}\tau \sim 2.3\mu\mathrm{m}$ for $V_{\mathrm{th}} = 2.3 \times 10^5$ m/s in silicon and $\tau \sim 10$ps as a rise time taken from electrical measurements in the literature[6,7]. This estimate gives a change in concentration of $1.26 \times 10^{18}$ cm$^{-3}$, which can be achieved in a SPAD with just a single initial electron resulting from the absorption of one photon.

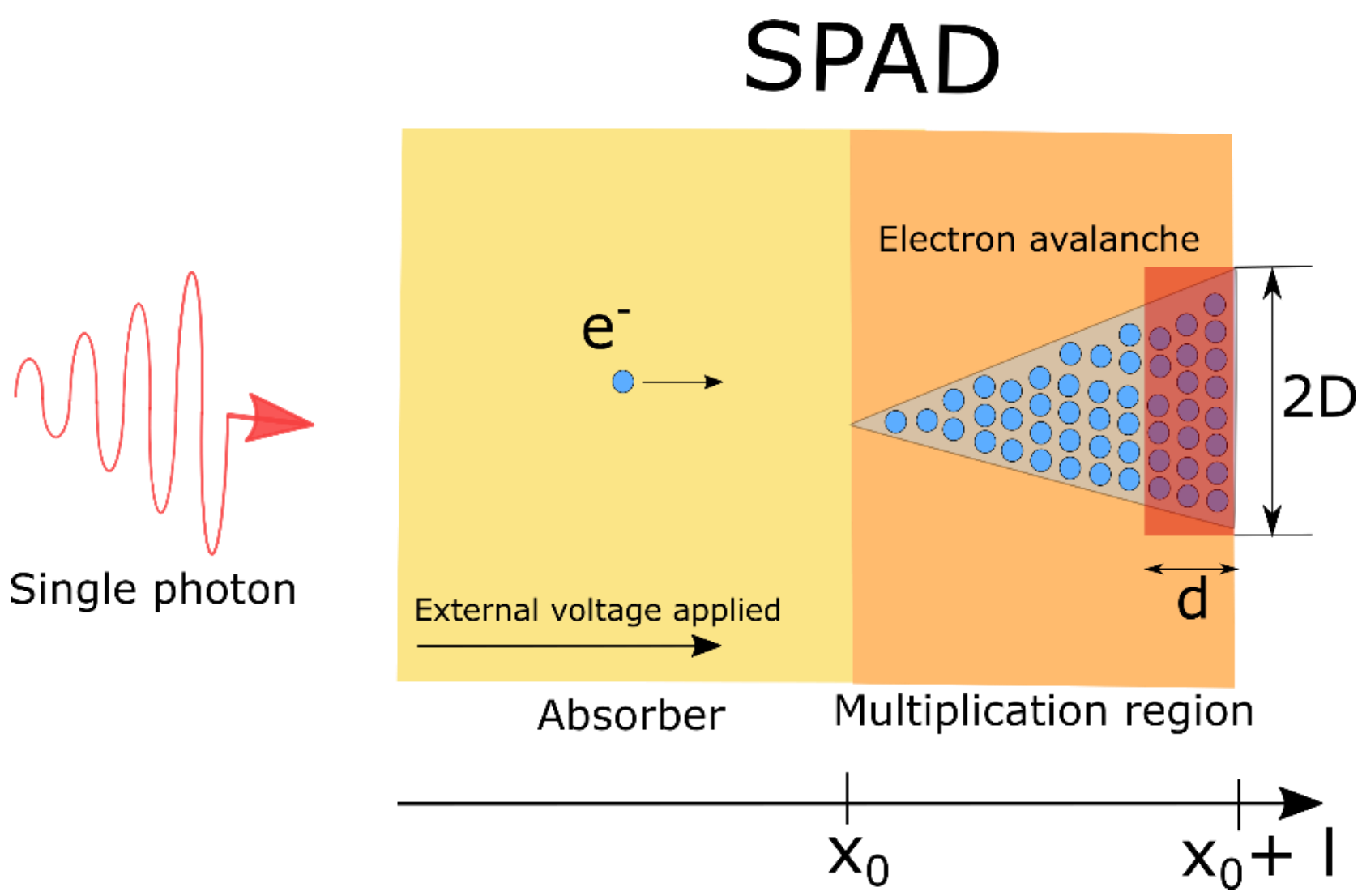

**Fig. S3: Schematics of SPAD.** Representation of electron avalanche in a one-dimensional model.

## S5.Amplitude modulation as a function of time delay

The dependence of the signal on the time delay between the pump and probe pulses is shown in **Fig. S4** and **Fig. 2a** of the main text. As discussed in the main text, two different effects contribute to the modulation. In this section, we are focusing on explanation of the "slow" effect only.

The amplitude dependence on the time delay between pulses (**Fig. 2a** and **Fig. S4**) contains two important features – the presence of a non-zero signal when the delay is negative and the absence of modulation in the range of delays from -17 to 17ns. As mentioned in the main text, the first feature can be explained by assuming that the probe pulse senses the modulation caused by the previous pump/control pulse. This effect implies that the relaxation time for the "slow" effect is longer than the repetition time between the subsequent measurements (the repetition rate is in the range 500 kHz - 5 MHz), which is confirmed by measurements of the relaxation time on a bigger time scale for delays (**Fig. 2b** in the main text). The second feature is thought to be due to the avalanches on a SPAD triggered by NIR probe pulses, which happen when the NIR power surpasses 15mW (**Fig. 4a** of the main text). In the case of small delay between pulses, the avalanche from the probe NIR beam prevents the generation of avalanches from the control beam. Similarly, the avalanche from a control beam blocks creation of avalanches from NIR probe beam if control pulse comes first. Thus, regardless of which pulse (the control/pump or the probe) initiated the avalanche, it does not allow the subsequent pulse to create another avalanche. This effectively results in the probe detected signal being the same for control "on" and "off" states. This in turn means that the amplitude of the AOM signal goes to zero in both negative and positive time delays around zero (within the dead time). Our findings confirm that he absence of the AOM signal lasts for about 34ns (from -17ns to 17ns), which roughly corresponds to twice of the dead time of the SPAD used in our measurements (16 ns, as stated in the datasheet for the Excelitas SPCM-AQRH-10-ND).

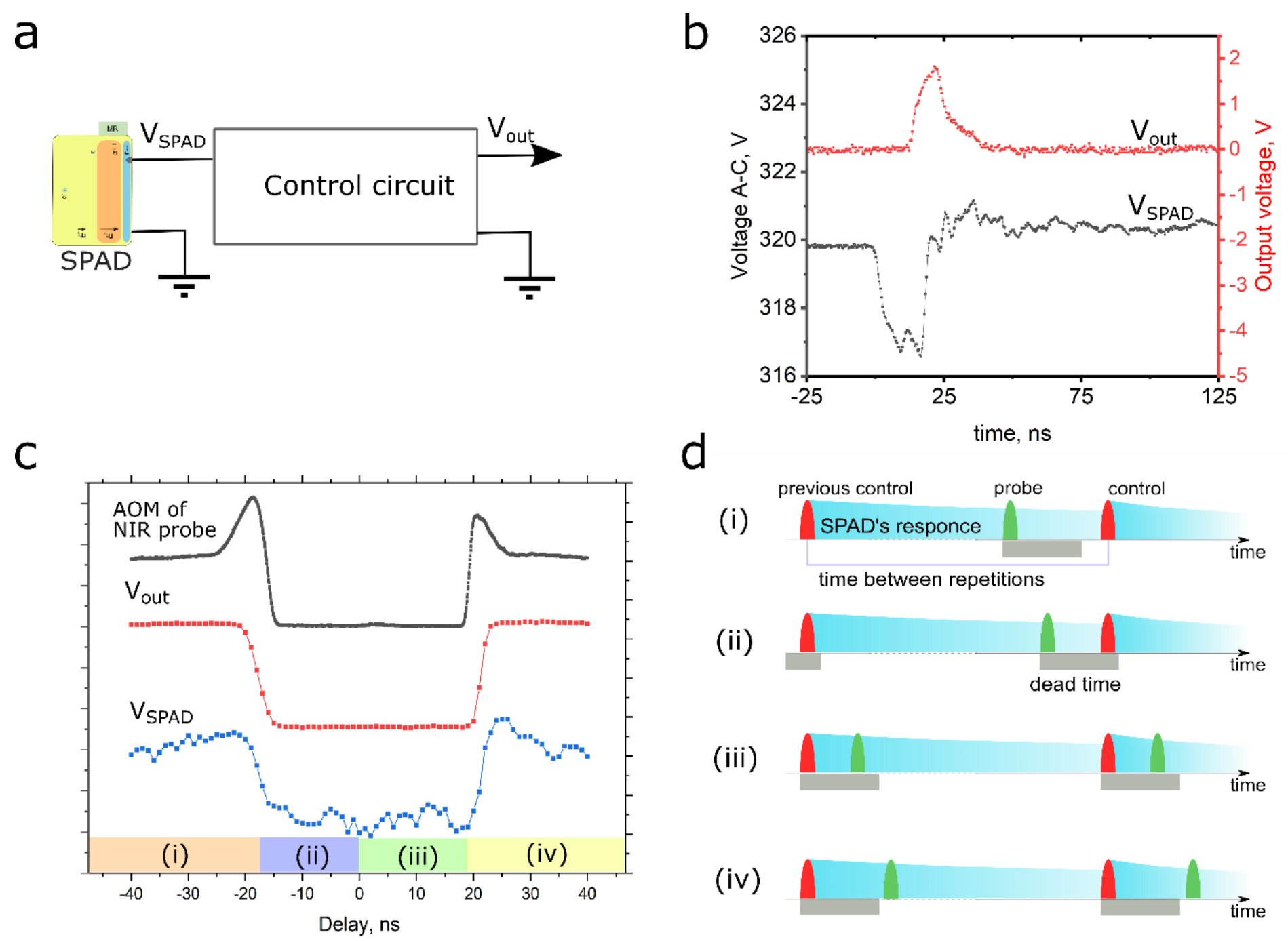

**Fig. S4: Electronic response of the SPAD. a,** Simplified schematics of a single-photon counting module (SPCM) consisting of a SPAD and controlling circuitry. **b,** Voltage as a function of time before($V_{\rm SPAD}$) and after($V_{\rm out}$) the controlling circuit board [8,9]. **c,** The amplitude of modulation of NIR probe beam (AOM) and integrated voltages as a function of delay between control and probe pulses. **d,** Pulse sequence for each case of time delay between pulses.

Using these two factors, we can build a model which describes the curve on the figure. In the case of the "slow" effect, the strength of modulation is proportional to the modulation of the electrical current flow through the SPAD, i.e., the number of triggered events, so that modulation amplitude $S \propto \Delta N = N_{on} - N_{off}$ , where $N_{on}$ and $N_{off}$ are triggering rates of a SPAD for chopper's ON and OFF states, respectively. This hypothesis is confirmed by the experimentally measured total count rate as a function of a delay between the pump and the probe pulses (**Fig.S4c**).

To describe the behavior of the curve in more detail, we divide it into 4 parts, as shown in **Fig. S4c**, where each part corresponds to a specific operating mode. For each case, we express the modulation of the triggering rates $\Delta N$ in terms of the probabilities of triggering an avalanche in the SPAD by the control or NIR probe pulses ($p_{vis}$ and $p_{NIR}$ respectively) and the repetition rate of the measurement (laser repetition rate) $R$, as described in the table below.

| Region of the graph in **Fig. S4c** | $N_{\rm ON}$ | $N_{\rm OFF}$ | $\Delta N$ |
|---|---|---|---|
| I | $(p_{\rm vis} + p_{\rm NIR}) \times R$ | $p_{\rm NIR} \times R$ | $p_{\rm vis} \times R$ |
| II | $(p_{\rm NIR} + (1 - p_{\rm NIR})p_{\rm vis}) \times R$ | $p_{\rm NIR} \times R$ | $(1 - p_{\rm NIR})p_{\rm vis} \times R$ |
| III | $(p_{\rm vis} + (1 - p_{\rm vis})p_{\rm NIR}) \times R$ | $p_{\rm NIR} \times R$ | $(1 - p_{\rm NIR})p_{\rm vis} \times R$ |
| IV | $(p_{\rm vis} + p_{\rm NIR}) \times R$ | $p_{\rm NIR} \times R$ | $p_{\rm vis} \times R$ |

In regions I and IV of the plot in **Fig. S4c**, the near-IR (probe) and visible (control) pulses are far away in time from each other compared to the SPAD's dead time so that they are detected independently. In these cases, the total count rate of the SPAD is simply the sum of the count rates from the pump and the probe pulses separately, so that $N_{\rm ON} = (p_{\rm vis} + p_{\rm NIR}) \times R$ . At the same time, the expression for count rate $N_{\rm OFF}$ when the chopper is in state OFF (i.e., the control beam is blocked) can be derived from $N_{\rm ON}$ by putting $p_{\rm vis} = 0$. So, in regions I and IV, the modulation of the count rate is proportional to the number of clicks triggered by the pump beam $\Delta N = N_{\mathrm{ON}} - N_{\mathrm{OFF}} = p_{\mathrm{vis}} \times R$, when the chopper is opened.

In regions II and III, pulses are blocking each other due to dead time of the SPAD, making formulas for $N_{\mathrm{ON}}$ more complicated. In these cases, each of these expressions is a summation of two possibilities. The first possibility is that an earlier pulse creates an event and blocks the triggering from the second pulse. The second possibility is that the first pulse is not creating an avalanche, making possible the detection of the subsequent pulse with a certain probability. The overall count rate in this case is a sum of these two possibilities $N_{\mathrm{ON}} = (p_{\mathrm{NIR}} + (1 - p_{\mathrm{NIR}})p_{\mathrm{vis}}) \times R$ or $N_{\mathrm{ON}} = (p_{\mathrm{vis}} + (1 - p_{\mathrm{vis}})p_{\mathrm{NIR}}) \times R$ , which is leading to the final expression for modulation of counts rate $\Delta N = (1 - p_{\mathrm{NIR}})p_{\mathrm{vis}} \times R$. As can be seen, it is decreasing with an increase of $p_{\mathrm{NIR}}$ and rising with increasing $p_{\mathrm{vis}}$ . Importantly, when $p_{\mathrm{NIR}} = 1$ the signal of modulation is equal to zero because $\Delta N = 0$, which corresponds to the case of relatively high NIR power, so that each pulse from the probe beam results in the triggering with 100% probability.

The presented simple model explains important features observed in the experiment. Specifically, $\Delta N$ in regions II and III is given by the same values regardless of which pulse is coming first. This prediction agrees with the experimental data (**Fig. S4c**). Another important conclusion from the model is that the signal should be dropping linearly when increasing the probability $p_{\mathrm{NIR}}$ for delays within the gap. This behavior is indeed observed and in accord with the experimental data in **Fig. 3a** of the main text. When the probe beam's power increases, the signal goes down, which is consistent with the prediction of the model. After the probe pulse probability reaches unity, the signal starts increasing because a different effect starts playing a dominant role under such conditions. Thus, both experimental curves are in good agreement with the behavior predicted by the model.

Despite consistency in explaining the observed gap-type behavior, this model doesn't explain the two peaks located right next to the limits of a gap (at 20 ns and -20 ns delays). We believe that a control circuitry board is responsible for such type of behavior[8–10]. To prove it, we performed a direct measurement of the voltage applied to a SPAD before feeding it into the control circuit board (**Fig. S4c**). The result shows similar behavior with two peaks, while the signal after the control circuit doesn't show such features. Those peaks are not as prominent compared to the data obtained for the optical modulation. Still, considering the highly nonlinear regime in the I-V curve of SPAD, this effect could lead to a significant increase in the electrical current. Thus, we believe that the peaks occur due to the circuit that applies additional voltage to a SPAD when two subsequent pulses are detected, with an approximate time separation of the dead time.

We also observed that the relative height of the peaks and their width can vary depending on the position of the SPAD, focusing depth, laser pulse duration, and other parameters. Furthermore, the presence or absence of a small peak corresponding to the "fast" effect is very sensitive to the SPAD position. All these interesting features are to be investigated in future studies.

## S6.COMSOL simulations

For a deeper understanding of the avalanche behavior, we numerically simulate carriers' density in a SPAD using the finite element method (Semiconductor Module, COMSOL Multiphysics v. 6.2). In the model[11], the classical current continuity equations are complemented with the generation (source) and recombination (sink) mechanisms for electrons or holes. We assume the empirical Okuto-Crowell model[12] for carrier generation rates $(R_i)$ due to impact ionization, producing an avalanche breakdown upon a high reverse electric bias [13]

$$\begin{pmatrix} \nabla \cdot \boldsymbol{J}_{\mathrm{n}} += -q_{\mathrm{e}} R_{\mathrm{n}} \\ \nabla \cdot \boldsymbol{J}_{\mathrm{p}} += q_{\mathrm{e}} R_{\mathrm{p}} \end{pmatrix},$$

$$R_i \stackrel{\mathrm{def}}{=} -\frac{1}{q_{\mathrm{e}}} \sum_{j=\mathrm{n,k}} \alpha_j |\boldsymbol{J}_j|, \alpha_i = a_i[1 + c_i(T - T_{ref})] E_{\parallel,i} \mathrm{e}^{-\left(\frac{b_i}{E_{\parallel,i}}[1+d_i(T-T_{\mathrm{ref}})]\right)^2}, (i = \{n, k\})$$

where, $\boldsymbol{J}_{\mathrm{n,p}}$ are the corresponding current densities; $q_{\mathrm{e}} = 1.6022 \times 10^{-19}\mathrm{C}$ is the elementary charge. The coefficients $\alpha_i$ define the impact ionization rates' dependence on the electric field $E_{\parallel,i}$ and the operation temperature $T$[12], $E_{\parallel,i}$ are the components of the *E*-field aligned with the electron or hole currents, and terms $T_{\mathrm{ref}}$, $a_i$, $b_i$, $c_i$, and $d_i$, are the material constants, taken from the embedded material library for Si. The recombination processes accounted for in the model include (i) the trap-assisted or Shockley– Read–Hall recombination, which is relevant for silicon (and other indirect band gap semiconductors) and (ii) the Auger recombination in which free carriers are involved in the process and become dominant at higher nonequilibrium carrier densities.

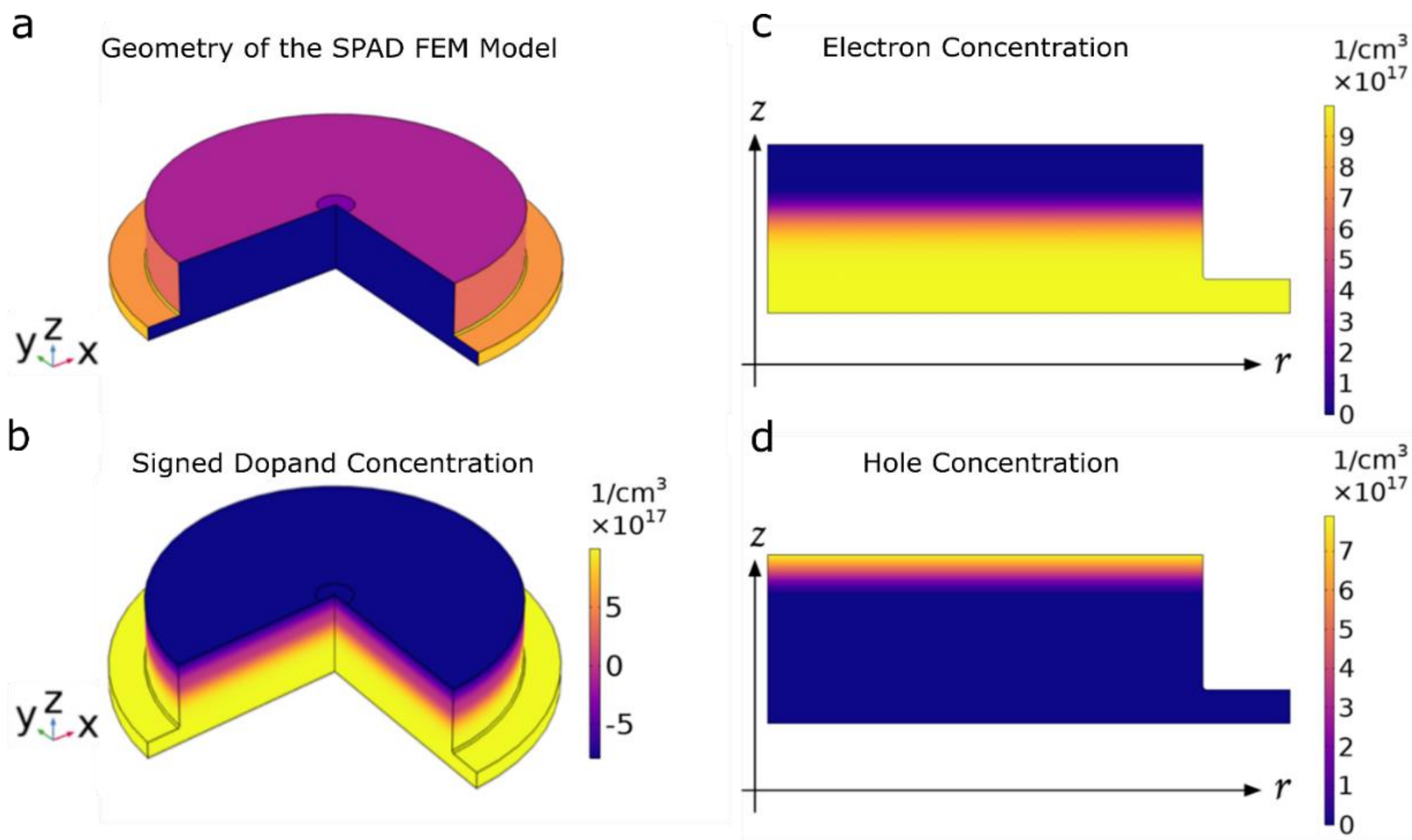

**Fig. S5: Device simulation of the SPAD under reverse bias**. **a,** The axisymmetric model FEM of the Si SPAD with a one-quarter cut-out; the dark-blue cross-section planes are used to depict the pseudo-color maps in panels **c** and **d**. In **a**, the anode is depicted as a dark purple circle at the center of the SPAD top face. The bottom orange ring is used as the cathode electrode adjacent to a solid n++ base. **b,** The gradient of the signed dopant concentration inside the device; two mirror-symmetric Gaussian distributions are assumed for the electrons and holes with decay lengths of $2\mu\mathrm{m}$. Electron **c** and hole **d** concentrations at t = 0, respectively.

The SPAD is modeled as a cylindrical structure with p-doped and n-doped regions with concentrations of dopants $N_{\mathrm{A0},}N_{\mathrm{D0}} = 10^{18}\,\mathrm{cm}^{-3}$(**Fig. S5**). The distribution of electrons and holes in the p-n junction is modeled as two mirror-symmetric Gaussian distributions with decay lengths of $2\mu\mathrm{m}$. The SPAD radius and height in the FEM model are, respectively, $r_{\mathrm{SPAD}} = 25\mu\mathrm{m}$, and its total height is $h_{\mathrm{SPAD}} = 12\mu\mathrm{m}$. The anode diameter is $d_{\mathrm{A}} = 5\mu\mathrm{m}$, the cathode ring width is $w_{\mathrm{C}} = 5\mu\mathrm{m}$, and the thickness of the uniformly doped base is $h_{\mathrm{C}} = 2\mu\mathrm{m}$. In contrast to a 2D geometry[11], we use an idealized 3D design, depicted in **Fig. S5**.

The avalanche breakdown is emulated by applying a reverse bias voltage ramped vs. time (see the red solid line in **Fig. S6a**). The ramp function has the amplitude of 11.4 V, the slope of 1000, and a relaxation offset of –50au; the starting time and duration of the smoothing transition zone are both equal to 100au. In **Fig. S6a**, the dashed lines depict the overlapping magnitudes of the dark currents vs. time (gray, anode; orange, cathode). Three moments, marked with blue circles in **Fig. S6a** present the pseudo-color frames of the impact ionization rates in panels (B, $t = 50\mathrm{au}$), (C, $t = 200\mathrm{au}$) and (D, $t = 400\mathrm{au}$).

Additional animations presenting the dynamics of the carriers (electrons and holes) are included as complementary Supporting materials files. In simulation, it is shown that the generated electron/hole density spreads in a small region within a short period compared to the overall lifetime of an avalanche. It is predicted that the generation of carriers occurs mainly within a very thin layer around the p-n junction of about 50nm in thickness, in accord with the number we obtained in S1. At the same time, it happens along the whole width of the sample in the lateral dimension, with the dominant contribution from a 1-2 µm region around the point where the avalanche starts.

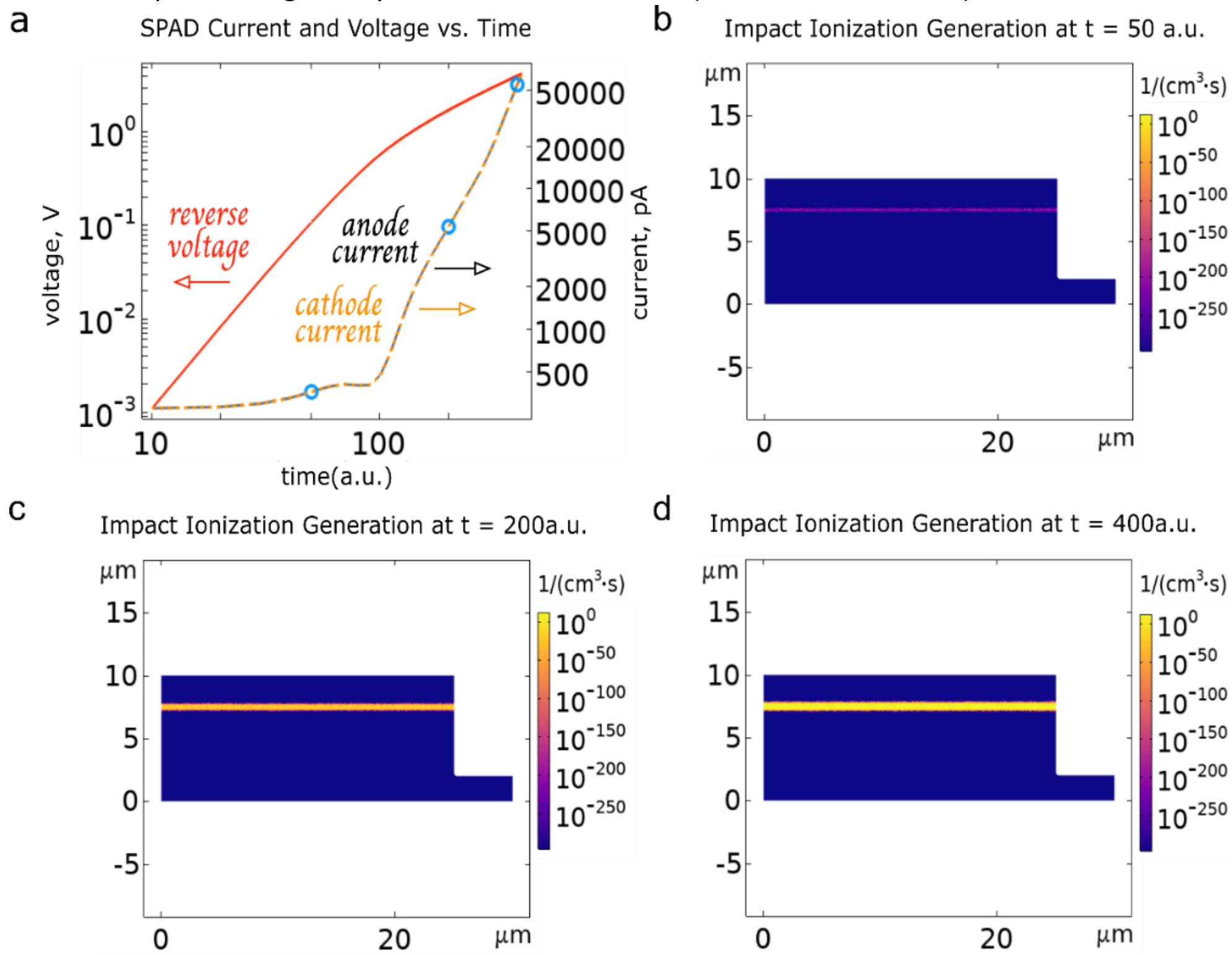

**Fig. S6: Reversed bias generation of the avalanche breakdown**. **a,** The reverse bias voltage ramped vs. time (red solid line, log-scale left vertical axis). The selected times ($t = \{50, 200, 400\} au$) are marked with blue circles at the overlapping dashed grey (anode) and orange (cathode) of the dark current magnitudes. **b, d,** The selected times are then used to depict the pseudo-color maps of the E-field dependent impact ionization rates in panels (**b**: $t = 50\text{au}$), (**c**: $t = 200\text{au}$), and (**d**: $t = 400\text{au}$).

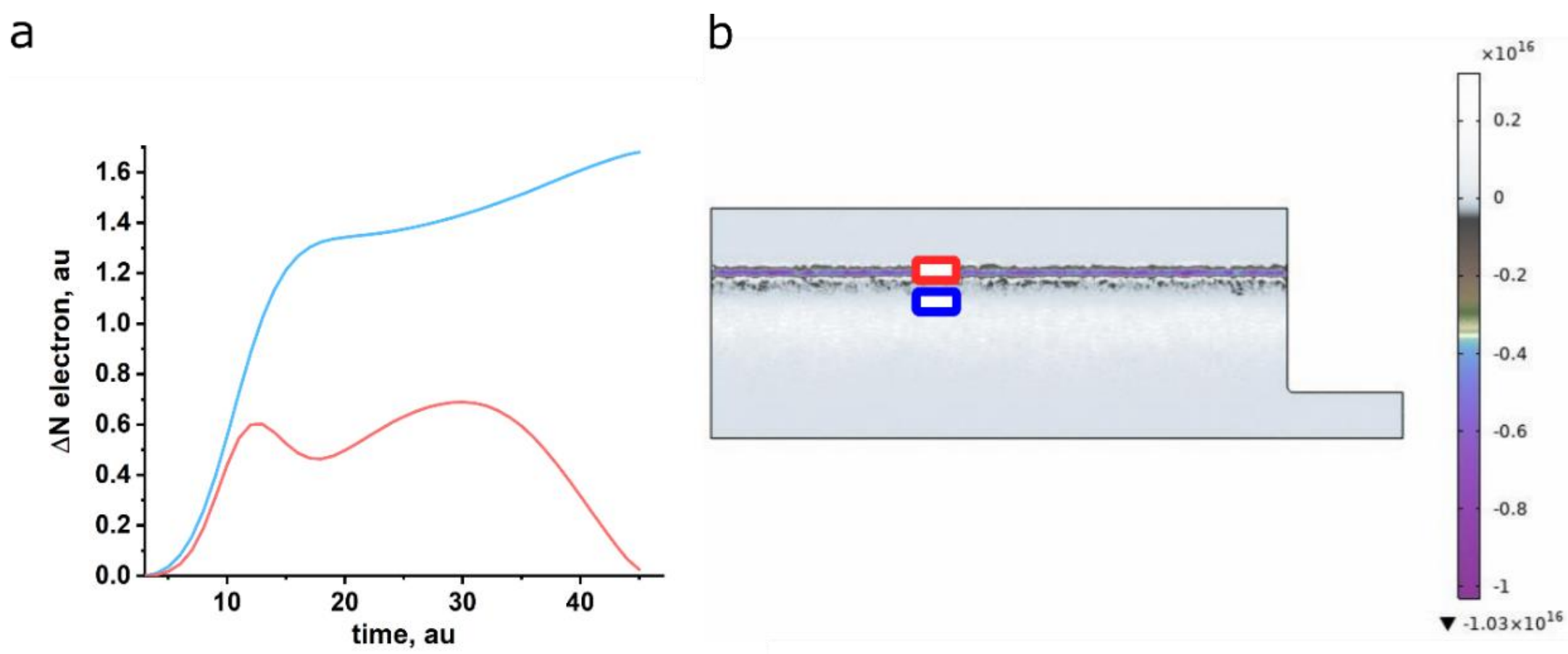

**Fig. S7: Simulations.** **a**, Time evolution of electron density in selected volumes of a SPAD (blue and red rectangles in **b**). **b,** The two-dimensional cross-section of differential electron density distribution $\Delta N_e$ at $t = 160\mathrm{au}$. The complete animated evolution is shown in Supplementary files.

The evolution of the charge density was extracted from simulations for the two spots in **Fig. 7b**. We calculate the injected electron density $\Delta N_e$ for the two closely spaced small volumes. The highest concentration is found for the volume labeled by the blue rectangle in **Fig S7b**. As shown in **Fig S7a**, in this region, the electron concentration rapidly grows up, reaching saturation at a certain moment. On the other hand, the electron concentration in the red rectangle (**Fig S7b**) demonstrates a fast decay almost to its initial concentration. This example demonstrates that the local charge dynamics can occur at a much faster rate compared to the overall avalanche time scale. This result can be used to reach significantly faster rates of modulation in future experiments. We note that the optimization of the SPAD performance at the device level goes beyond the scope of this paper and it will be discussed elsewhere; in this study we simply employed a commercial SPAD for our experiments.

## S7.Estimations of sensitivity

The experiments that were conducted demonstrate the sensitivity to the optical control beam, consisting, on average, of substantially less than one photon per pulse, i.e., the system should be able to capture single-photon events reliably. However, the measurement presented in the paper is limited by the signal-to-noise ratio (SNR), making it impossible to capture single-shot events, which are interesting for some practical applications. To estimate the sensitivity of our proof-of-principle realization, we measured the modulation of the NIR probe signal directly on the oscilloscope (**Fig. S8**). In this experiment, we observed the minimal time of switching between the ON and OFF states of a chopper was around 0.15 ms, which is limited by the rotational speed of a mechanical chopper. In this case, the upper estimation of the minimal number of photons required to switch the system between ON and OFF states is only $600$ photons ($0.15\mathrm{ms} \times 80\mathrm{MHz} \times 0.05$ photons/pulse). However, due to low SNR, it cannot be detected through a single measurement, making it necessary to integrate over several subsequent measurements. To estimate the minimal number of measurements required to see switching, we acquired several data sets with different averaging (**Fig. S8**). As expected, the observed modulation of the NIR probe between ON and OFF chopper states becomes more prominent with more statistical data due to the increased signal-to-noise ratio. As a criterion for the limit at which point 'ON/OFF' states can be considered distinguishable, we introduce a contrast $C = \frac{\langle N_{\mathrm{OFF}} \rangle - \langle N_{\mathrm{ON}} \rangle}{\sqrt{\langle N_{\mathrm{OFF}}^2 \rangle + \langle N_{\mathrm{ON}}^2 \rangle}}$ , which is the difference between mean values for ON/OFF state divided by the sum of the respective standard deviations. We assume that contrast $C > 0.5$ is necessary for measuring distinguishable levels from a single curve corresponding to the case when the modulation amplitude is comparable with the noise. This threshold is achieved after averaging approximately over 50 data sets (**Fig. S8f**). This averaging is equivalent to a single shot switching with $50 \times 600 = 30{,}000$ photons, if it was possible to scale the effect by increasing the number of photons in the control pulse. Reducing the shot noise and increasing SNR could be realized in two different ways – by increasing the power of the probe beam or by introducing photonic cavities, which we plan to do in our future research. One should keep in mind that higher powers for 1550 nm wavelength will lead to a higher dark count rate on a silicon SPAD, which could be avoided by using a longer wavelength of the probe beam.

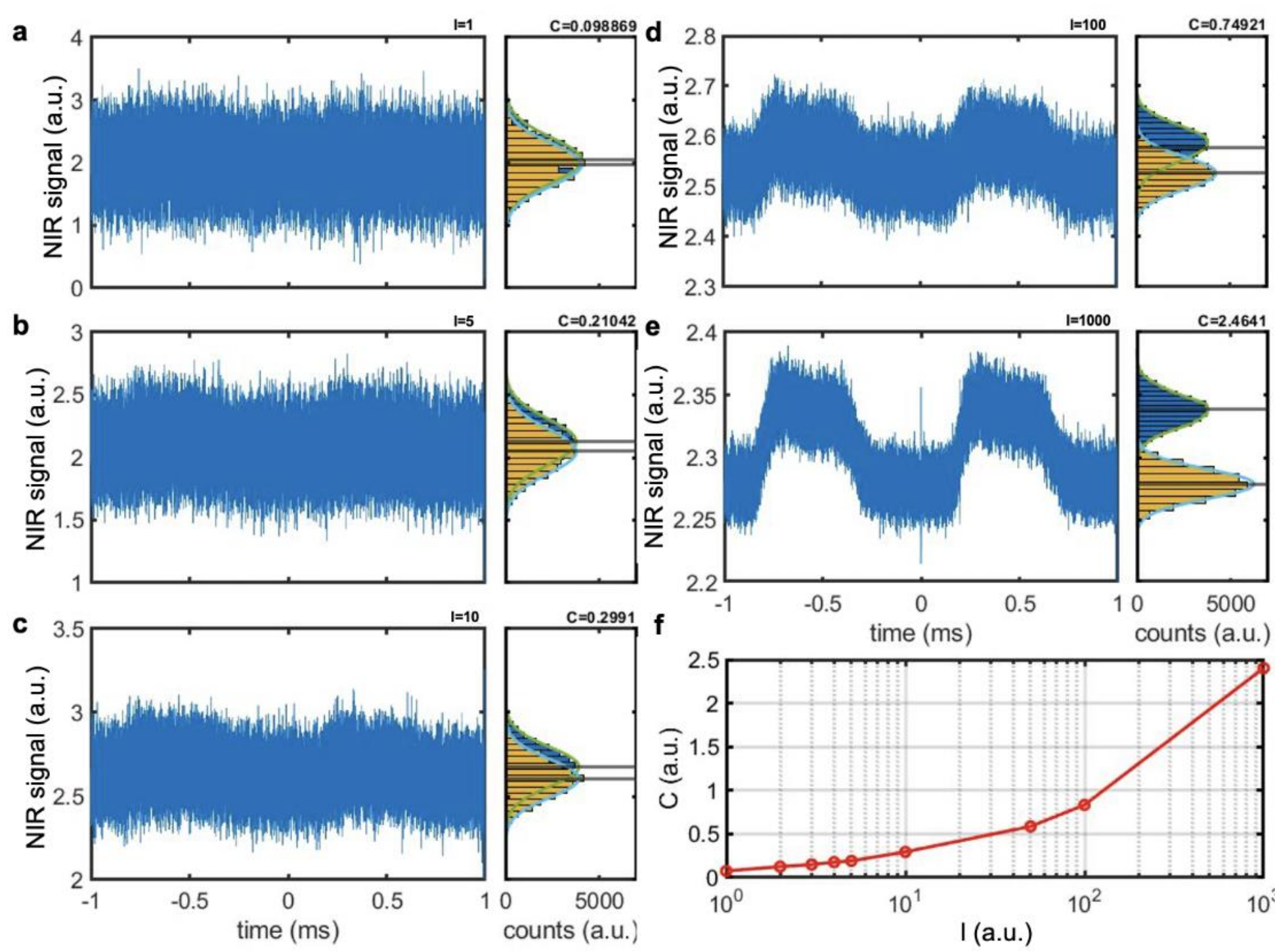

**Fig. S8: Probe signal and modulation contrast at different levels of averaging a-e** Signal plots of NIR probe beam intensity averaged over a different number of data sets $l = 1; 5; 10; 100; 1000$ for $0.05$ average number of photons per pulse. The rotated histograms on the right side of each signal plot analyze the probe's signals for ON/OFF (yellow/blue) chopper's states. **f,** The contrast as a function of the averaging number. The dot markers indicate the measured data; a solid curve is a guide for an eye.

## S8.Summary tables

| *Material/work* | $n_2, m^2 W^{-1}$ | *Probe wavelength, nm* |
|---|---|---|
| **Silicon SPAD/this work** | $\mathbf{1.3 \times 10^{-2}}$ | **1550** |
| Gold[14,15] | $2.6 \times 10^{-14}$ | 532 |
| AZO[16] | $5.2 \times 10^{-16}$ | 1311 |
| ITO[17,18] | $1.1 \times 10^{-14}$ | 1240 |
| TDBC[19] | $1.7 \times 10^{-14}$ | 500 |
| HTJSq[19] | $3.5 \times 10^{-15}$ | 564 |
| Silicon[20] | $5 \times 10^{-18}$ | 1500 |
| Fiber[21,22] | $2.9 \times 10^{-20}$ | 1500 |
| CS[23] | $2.3 \times 10^{-14}$ | N/A |
| Cold atoms(EIT, BEC)[24] | $2.0 \times 10^{-5}$ | 589 |
| Fluorescein dye in glass[25] | $3.5 \times 10^{-7}$ | 488 |
| Polymer PTS[26] | $-2.0 \times 10^{-14}$ | 650-700 |
| LBO[27] | $0.26 \times 10^{-19}$ | 780 |
| Lithium Niobate[28] | $2.5 \times 10^{-19}$ | 532 |
| Atomic Rb [29,30] | $\sim 10^{-10}$ | 780 |

**Table S1:** Comparison of nonlinear refractive index for different materials

| Figure | Fitting function | Fitting parameters |
|---|---|---|
| Figure 2b | $Ae^{-\frac{t}{\tau}}$ | $A = 1.25 \times 10^{-5}$<br>$\tau = 13.46\ \mu s$ |
| Figure 2d | $A\left[1 + Tanh\left(-\frac{t-t_0}{\tau_1}\right)\right] e^{-\frac{t-t_0}{\tau_2}} + B$ | $A = 3.5 \times 10^{-6}$<br>$B = 1.75 \times 10^{-6}$<br>$\tau_1 = 1.12\text{ns}$<br>$\tau_2 = 2.15\text{ns}$<br>$t_0 = 2.02\text{ns}$ |
| Figure 3b | $Ax + B$ | $A = 6.1 \times 10^{-4}$<br>$B = -2 \times 10^{-5}$ |
| Figure 3d | $Ax + B$ | $A = 2.4 \times 10^{-2}$<br>$B = -3.3 \times 10^{-4}$ |

**Table S2:** Fitting functions and their parameters for plots in the paper